# TRANSVERSE INSTABILITIES OF COASTING BEAMS WITH SPACE CHARGE*

A. Burov, V. Lebedev

FNAL, Batavia, IL 60510, U.S.A.

*Abstract*

Transverse beam stability is strongly affected by the beam space charge. Usually it is analyzed with the rigid-beam model. However this model is only valid when a bare (not affected by the space charge) tune spread is small compared to the space charge tune shift. This condition specifies a relatively small area of parameters which, however, is the most interesting for practical applications. The Landau damping rate and the beam Schottky spectra are computed assuming that validity condition is satisfied. The results are applied to a round Gaussian beam. The stability thresholds are described by simple fits for the cases of chromatic and octupole tune spreads.

## INTRODUCTION

Particle interaction via the walls of the vacuum chamber is conventionally described by the wake functions and impedances. In absence of damping, this interaction leads to beam coherent instabilities. However, if there are particles in resonance with coherent motion, they effectively exchange their incoherent energy with the energy of coherent oscillations. If the phase space density of the resonant particles is sufficiently large, the instability is stabilized by this mechanism, called the Landau damping. Contrary to the wake fields, Coulomb interaction does not drive the instability by itself, since it preserves the total energy and momentum. However, the collective Coulomb field can strongly affect the beam stability because it separates coherent and incoherent frequencies. Indeed, when the beam oscillates as a whole, its collective motion does not see the space charge, while an individual particle does. Thus, if the coherent and incoherent frequencies are separated, there are no resonant particles, and no Landau damping.

To analyze the beam stability with space charge, an effective method was suggested by D. Möhl and H. Schönauer in 1974 [1]. To describe transverse oscillations of a coasting beam, they introduced a linearized equation of motion:

$$\frac{d^2 x_i}{dt^2} + \Omega_i^2 Q_i^2 x_i + 2\Omega_0^2 Q_0 \left(\Delta Q_c \bar{x} + \Delta Q_{sc}(x_i - \bar{x})\right) = 0. \quad (1)$$

Here $x_i$ is the offset of $i$-th particle, $\Omega_i$, $Q_i$, $\Delta Q_{sc}$ are its revolution frequency, the tune and the direct space charge tune shift, $\Omega_0$, $Q_0$ are the average revolution frequency and tune, $\bar{x}$ is the offset of beam center and $\Delta Q_c$ is the impedance-driven coherent tune shift. Although perturbation of a particle motion depends on its amplitudes, this equation assumes that the beam oscillates as a rigid body when the coherent beam fields are computed. Consequently, the beam coherent motion is completely described by the dipole offset $\bar{x}$. This assumption is correct if all lattice frequencies $\Omega_i Q_i$ are identical. In this case, all particles respond identically to the coherent field, $\delta x_i = \bar{x}$, consequently, the beam oscillates as a rigid body, and the spread of the space charge tune shifts does not matter. However, a spread of the lattice frequencies generally makes the rigid-body model of Eq. (1) incorrect. Indeed, an individual response to the coherent field is determined by the separation of the individual lattice frequency from the coherent frequency, which varies from particle to particle. Since individual responses are not identical, the beam shape is not preserved in the dipole oscillations, so the rigid-body model of Eq. (1) is not self-consistent and generally cannot be justified.

In 2001, M. Blaskiewicz showed a way to analyze the problem, avoiding the rigid-beam assumption [2]. Within a one-dimensional model, he developed an integral equation on the phase space density perturbation. He found two cases when his equation gives the same result as the rigid-beam approach. The first case was the Lorentz momentum distribution, and the second one was the water-bag distribution over the transverse actions. With some additional model simplifications, he plotted several stability diagrams for distributions close to Gaussian. The same problem of self-consistent beam stability analysis was recently examined by D. Pestrikov [3]. Considering a two-dimensional model, he came to a general integral equation and found it "too complicated even for a numerical solving." To proceed, he considered a single-dimensional problem, came to the same integral equation as M. Blaskiewicz, and reproduced his Lorentz and water-bag results. For a Gaussian distribution, he plotted additional stability diagrams, and found no anti-damping, found earlier in his rigid-beam model studies [4]. Indeed, Landau anti-damping cannot exist at all if the distribution is close to Gaussian: this is a mere consequence of the second law of thermodynamics. A Hamiltonian system in thermal equilibrium is always stable. Appearance of Landau anti-damping in the rigid-beam model is a striking example of how wrong the results of this model can be. Rigid-beam stability diagrams were presented in several

___
* Work supported by the U.S. Department of Energy under contract No. DE-AC02-76CH03000

papers [4-6]; however, the range of their applicability was not clarified.

## MODEL JUSTIFICATION

As mentioned above, the rigid-beam model is correct if all the lattice frequencies are identical, $\Omega_i Q_i = \Omega_0 Q_0$. This case is simple, but not so interesting, since there is no Landau damping, and any impedance with non-zero real part makes the beam unstable. Now let us assume that the lattice frequency spread is sufficiently small so that the rigid-beam model would still be a good approximation. That requires the rms spread of the lattice frequencies $\sigma(\Omega_i Q_i)$ to be small compared with the separation frequency, which is a difference of the coherent frequency from an average incoherent one:

$$\sigma(\Omega_i Q_i) << |\Delta\Omega_{sep}| \equiv \Omega_0 |\text{Re}(\Delta Q_c) - \langle \Delta Q_{sc} \rangle| \;, \quad (2)$$

where $\langle \Delta Q_{sc} \rangle$ is the average space charge tune shift. In this case, the rigid-beam model is still a good approximation; but there is small amount of the resonant particles in tails of the distribution, yielding small Landau damping. If the impedance-driven instability rate $\Omega_0 \text{Im}(\Delta Q_c)$ is also small, even this tiny amount of Landau damping can be sufficient for the beam stabilization. Thus, when the instability rate is much smaller than the separation frequency, or

$$\text{Im}(\Delta Q_c) << |\text{Re}(\Delta Q_c) - \langle \Delta Q_{sc} \rangle| \;, \quad (3)$$

a relatively small frequency spread is sufficient to stabilize the beam. Near the threshold, the small frequency spread is not significant for a bulk of the beam, which oscillates almost the same way as for zero tune spread. The tiny amount of the resonant particles has almost no influence on the coherent motion, except a slow transfer of the coherent energy into incoherent one, and thus, a slow collective mode damping. In other words, when Eq. (3) is satisfied, the rigid-beam model is applicable for calculation of Landau damping required for beam stabilization. This energy-based calculation of Landau damping leads to the same result as a formal solution of the dispersion equation [8].

In this paper, we limit ourselves to a case of thin tail, or small frequency spread approximation of Eq. (2), where the rigid-beam model is applicable. This allows us to calculate the Landau damping and the threshold parameters of the beam for a relatively small growth rate (3). Our primary interest is the threshold calculation. This is additionally simplified due to exponentially small phase space density of resonant particles, and consequently the Landau damping. When the damping rate is a steep function of the dimensionless frequency separation $\Delta\Omega_{sep}/\sigma(\Omega_i Q_i) >> 1$, the threshold condition mostly determines this big ratio, being only slightly dependent on the coherent growth rate.

In practice, the far tails of the distributions are not well-measured or well-reproducible, so that even exact formulas cannot produce reliable results for the instability growth rates. On the contrary, the stability threshold for the beam intensity, being an inverse function of the Landau damping rate, depends much weaker on specific behavior of the distribution tails, and therefore can be predicted much better. Note also that the condition of small growth rate of Eq. (3) is typically well-satisfied for low and medium energy hadron machines, mainly addressed by this paper.

## DISPERSION EQUATION

After validity limits of the rigid-beam model are specified, a solution of Eq. (1) can be considered in more details. Assuming $x_i(t) \propto \exp(-i\omega t)$ and $\omega \equiv \Omega_0(n + Q_0 + \nu)$, one obtains the dispersion relation for the eigenvalue $\nu$ [1]:

$$\varepsilon(\nu) \equiv 1 - \int \frac{(\Delta Q_c(\omega) - \Delta Q_{sc}(J_x, J_y)) f_x J_x d\Gamma}{\Delta Q_l(J_x, J_y, \hat{p}) + \Delta Q_{sc}(J_x, J_y) - \nu - i0} = 0;$$

$$d\Gamma \equiv dJ_x dJ_y d\hat{p}; \quad (4)$$

$$\nu = \omega/\Omega_0 - (n + Q_0).$$

$$f = f(J_x, J_y, \hat{p}); \quad f_x \equiv \frac{\partial f}{\partial J_x}; \quad \int f dJ_x dJ_y d\hat{p} = 1.$$

Here, all the notations are rather conventional: $J_x$ and $J_y$ are the transverse actions; $\hat{p} = \Delta p/p$ is a relative momentum offset; $\Delta Q_l(J_x, J_y, \hat{p})$ is the total lattice-related tune shift; $\Delta Q_{sc}(J_x, J_y)$ is the direct space charge tune shift as a function of the amplitudes; and $\eta = 1/\gamma_t^2 - 1/\gamma^2$ is the slippage factor. The transverse actions are normalized by the beam rms emittances $x = \sqrt{2\varepsilon_x J_x \beta_x} \cos\psi_x$ and similarly for $y$. That results in that

$$\int f J_{x,y} dJ_x dJ_y d\hat{p} = 1 \;.$$

In this paper we will only consider the first two terms contributing to the total lattice-related tune shift: the chromatic contribution and the contribution due to octupole non-linearity so that: $\Delta Q_l(J_x, J_y, \hat{p}) = (\xi - (n+Q)\eta)\hat{p} + \Delta Q_o(J_x, J_y); \; n = 0, \pm 1, \pm 2,... \;.$ The coherent shift $\Delta Q_c(\omega)$ describes the beam interaction with the wall. This interaction produces both the dipole and quadrupole forces, or, in other words, driving and detuning wakes [9]. Thus, the entire force acting on $i$-th particle can be expressed as $F_i = W\bar{x} + Dx_i$, with $W$ as the conventional dipole (or driving) wake function, and $D$ as the quadrupole (or detuning) wake function. By definition, only the driving wake term contributes into the coherent shift $\Delta Q_c(\omega) \propto W$. For the coasting beam, the detuning wake simply shifts all tunes by the same amount. It makes no change for stability analysis and therefore is omitted below.

A conventional method of analysis of the dispersion equation results in a stability diagram in the complex plane of the coherent shift $\Delta Q_c(\omega)$. However, if the rigid beam model is used, only computations of tails of the diagram result in a reliable answer because inequalities of Eqs. (2) and (3) are not fulfilled for the diagram's main part. In the area of tails, however, another significant step can be done: the rate of Landau damping can be calculated and expressed in terms of a regular integral of the distribution function $f$.

## LANDAU DAMPING

When the condition (2) is satisfied, the rate of Landau damping can be found from Eq. (4). Note that this dispersion equation formally defines the dielectric function $\varepsilon(\nu)$ for values of $\nu$ located on the real axis and the upper half-plane, $\text{Im}(\nu) \geq 0$. To obtain it in the lower half-plane, where roots of the dispersion equation are located, the direct use of Eq. (4) is invalid; instead, a complex extension of the analytical function $\varepsilon(\nu)$ has to be used. This can be done in the following way. First, let the eigenvalue $\nu$ be real, $\text{Im}\,\nu = 0$, and solve the dispersion equation for the coherent shift as a function of the eigenvalue. Then the imaginary part of the found coherent shift $\text{Im}\,\Delta Q_c$ is equal to the Landau damping $\Lambda$, since at the threshold, $\text{Im}\,\nu = 0$, they exactly compensate each other. After expansion of the integral denominators over a small relative tune spread $\Delta Q_l / \Delta Q_{sep}$ for $\text{Re}\,\Delta Q_c$, the result for the eigenvalue $\nu = \nu_c$ and the damping rate $\Lambda$ follows (see the Appendix):

$$\nu_c = \text{Re}\,\Delta Q_c + \delta Q^{(1)} + \delta Q^{(2)},$$
$$\Lambda = -\pi \langle \Delta Q_{sep} \rangle \int \Delta Q_{sep} f_x J_x \, \delta(\Delta Q_l + \Delta Q_{sc} - \nu_c) d\Gamma, \quad (5)$$
$$\Delta Q_{sep} \equiv \text{Re}\,\Delta Q_c - \Delta Q_{sc}(J_x, J_y), \quad \langle \Delta Q_{sep} \rangle \equiv -\left( \int \frac{f_x J_x d\Gamma}{\Delta Q_{sep}} \right)^{-1},$$
$$\delta Q^{(1)} = -\langle \Delta Q_{sep} \rangle \int \frac{\Delta Q_l f_x J_x d\Gamma}{\Delta Q_{sep}}.$$
$$\delta Q^{(2)} = -\langle \Delta Q_{sep} \rangle \int \frac{\Delta Q_l^2 f_x J_x d\Gamma}{\Delta Q_{sep}^2}.$$

Note that the sign of the damping rate $\Lambda$ is always determined by the sign of the derivative of the distribution function $f_x = \partial f / \partial J_x$ for the resonance particles, similar to the classical Landau result for the plasma oscillations (no antidamping for monotonic distributions). Note also that corrections $\delta Q^{(1)}$ and $\delta Q^{(2)}$ to the real part of the eigenvalue play a role when the distribution function drops exponentially; in this case even a small correction to the tune of the resonant particles significantly affects the damping rate $\Lambda$.

Let us first assume that the tune spread is purely chromatic. In this case the first-order correction is equal to zero, $\delta Q^{(1)} = 0$, and only the second-order correction remains, $\delta Q^{(2)}$. For the Gaussian momentum distribution, $f \propto \exp(-\hat{p}^2 / 2\sigma_p^2)$, and constant transverse density, $\Delta Q_{sc} = \text{const}$, the second-order correction is determined by the following equation: $\delta Q^{(2)} = \sigma_{vp}^2 / \Delta Q_{sep}$, where $\sigma_{vp} \equiv |\xi - (n+Q)\eta|\sigma_p$. That yields the damping rate:

$$\Lambda = \sqrt{\frac{\pi}{2}} \frac{\Delta Q_{sep}^2}{\sigma_{vp}} \exp\left(-\frac{\Delta Q_{sep}^2}{2\sigma_{vp}^2} - 1\right). \quad (6)$$

Note that this result is $e$ times smaller than a simple-minded formula neglecting the second-order term $\delta Q^{(2)}$.

Another possibility to stabilize the beam is an introduction of octupole non-linearity. Contrary to the chromatic spread, the first-order correction to the eigenvalue $\delta Q^{(1)}$ is non-zero here. For the Gaussian transverse distribution, including this first-order correction reduces the rate $\Lambda$ by a constant factor ~2-3, similar to the role of the second-order term for the chromatic tune spread. For the octupole tune spread the second-order term makes only a small correction to the damping rate and can be neglected.

As was pointed out above, the rigid-beam model is valid only if the frequency spread is small compared with the separation frequency, $\delta Q^{(1,2)} / \langle \Delta Q_{sep} \rangle \ll 1$. Accounting the tune corrections $\delta Q^{(1)}, \delta Q^{(2)}$ within the rigid-beam approximation assumes that the inaccuracy of the model is smaller than these corrections. The correctness of this assumption is a subject of separate study. Presently, we can only refer to a specific example of chromatic tune spread for a Gaussian beam, considered in Ref. [3] within a framework of one-dimensional self-consistent model, compared with the rigid-beam result. As it is clearly seen from a presented stability diagram, the discrepancy between the two results is rather small, ~ 10-20% in the area of rigid-beam model validity. This suggests that accounting the eigenvalue corrections $\delta Q^{(1)}, \delta Q^{(2)}$ is within the model accuracy, and thus it is legitimate. Finally, it should be noted that although the corrections $\delta Q^{(1)}, \delta Q^{(2)}$ change the damping rate $\Lambda$ by 2-3 times, their influence on the threshold space charge over the tune spread value is relatively small, since the Landau damping exponentially depends on beam parameters (like in Eq. (6)), and an error in the pre-exponential factor (~2-3) only slightly modifies the threshold.

## THRESHOLD LINES

As it was stated above, rigid-beam stability diagrams are mostly invalid if the space charge is present. A small correct part of them lies typically so close to zero that it is hard to resolve details on the pictures usually presented in the literature (Ref. [4-6]). Therefore we do not draw these diagrams here and present the stability threshold in a different way. Indeed, Eq. (6) shows that the stability condition depends on two dimensionless parameters. The first parameter determines to what extent the coherent and incoherent frequencies are separated; obviously, it is defined by the ratio of the separation frequency over the lattice frequency spread. The second parameter shows how strong is the instability to be suppressed by the Landau damping; it can be described by the coherent growth rate $\Omega_0 \operatorname{Im}\Delta Q_c$ in units of the separation frequency. A dependence of the threshold dimensionless separation over dimensionless coherent growth can be called the threshold line. In this section we present it for round Gaussian beams. The problem is solved both for a pure chromatic tune spread,

$$\Delta Q_l = (\xi - (n+Q)\eta)\hat{p} \equiv \sigma_{vp}\hat{p}/\sigma_p,$$

and for an axially-symmetric octupole-induced spread,

$$\Delta Q_l = \Delta\sigma_{vo}(J_x + J_y)/2 > 0.$$

The results are presented in Figures 1, 2. Here we additionally assume that $|\operatorname{Re}\Delta Q_c| \ll |\langle\Delta Q_{sc}\rangle|$. We do not consider negative sign of the octupoles, since they would detune even more incoherent frequencies from the coherent line, making the beam more unstable.

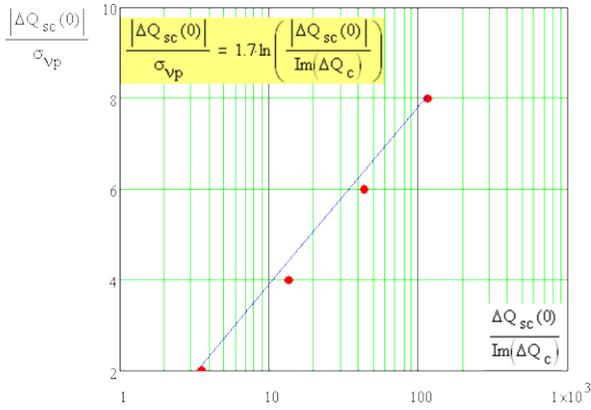

Figure 1: Threshold line for the chromatic tune spread. The dimensionless maximal space charge tune shift $|\Delta Q_{sc}(0)|/\sigma_{vp}$ is plotted versus dimensionless growth time $|\Delta Q_{sc}(0)|/\operatorname{Im}\Delta Q_c$. The dots are numerical results, and the line is a fit with the formula highlighted in yellow.

Above we used a following presentation for the space charge tune shift of a round Gaussian beam as a function of the transverse actions $J_x, J_y$ [10]:

$$\Delta Q_{sc}(J_x, J_y) = \Delta Q_{sc}(0)\int_0^1 \frac{\left[\mathrm{I}_0\!\left(\frac{J_x z}{2}\right) - \mathrm{I}_1\!\left(\frac{J_x z}{2}\right)\right]\mathrm{I}_0\!\left(\frac{J_y z}{2}\right)}{\exp(z(J_x + J_y)/2)} dz \,. \quad (7)$$

Here $\Delta Q_{sc}(0) = -r_p\lambda C/(4\pi\beta^2\gamma^3\varepsilon)$ is the maximal space charge tune shift with $\lambda$ as the linear density, $C$ as the orbit circumference, $r_p$ as the classical radius of the beam particles, $\varepsilon$ as the unnormalized rms emittance, and $\beta, \gamma$ as relativistic factors. For numerical calculations, we approximated the exact result (7) by the following fit:

$$\Delta Q_{sc}(a_x, a_y) = \Delta Q_{sc}(0)\frac{192 - 11a_x - 18\sqrt{a_x a_y} + 3a_y^2}{192 - 11a_x - 18\sqrt{a_x a_y} + 3a_y^2 + 36a_x^2 + 24a_y^2} \,, \quad (8)$$

$$a_{x,y} \equiv \sqrt{2J_{x,y}}$$

which is accurate within a few percent for $a_x, a_y \leq 6$; it has the right Tailor expansion at small amplitudes and the right asymptotic behavior at large amplitudes.

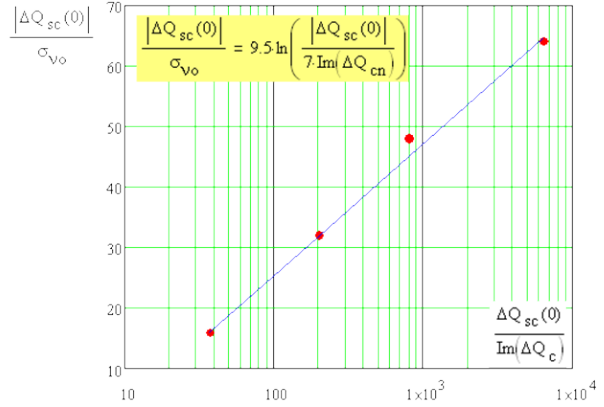

Figure 2: Threshold line for the octupole tune spread. Notations are similar to Fig.1.

As it is seen from the plots, the suggested fits for the threshold lines

$$\frac{|\Delta Q_{sc}(0)|}{\sigma_{vp}} = 1.7\ln\!\left(\frac{|\Delta Q_{sc}(0)|}{\operatorname{Im}\Delta Q_c}\right) \quad (9)$$

for the chromatic spread, and

$$\frac{|\Delta Q_{sc}(0)|}{\sigma_{vo}} = 9.5\ln\!\left(\frac{|\Delta Q_{sc}(0)|}{7\operatorname{Im}\Delta Q_c}\right) \quad (10)$$

for the symmetric octupole spread are accurate within 10% or better.

Note that the stabilizing rms tune spread is 3-4 times smaller for the octupoles than for the chromatic case. The reason is that the octupole-driven tune shift goes quadratically with amplitudes, while the chromatic tune shift is a linear function of the momentum offset.

It is instructive to compare thresholds for dimensionless growth rates $\operatorname{Im}\Delta Q_c/\Delta Q_{sep}$ of the Gaussian beam (9) with $|\Delta Q_{sc}(0)| \cong 2\Delta Q_{sep}$, and the KV beam (6) with $\Lambda = \operatorname{Im}(\Delta Q_c)$. The two thresholds are close at $\Delta Q_{sep}/\sigma_{vp} = 2$, but they diverge exponentially with increase of the separation frequency. At $\Delta Q_{sep}/\sigma_{vp} = 5$, the Gaussian threshold growth rate is almost 3 orders of magnitude higher than that for the KV beam. The reason for this advantage of the Gaussian beam is that its resonant particles may have not so high momentum offset in expense of higher transverse amplitudes, where the space charge tune shift goes down. For the KV beam this is impossible, and that is why the KV threshold growth rate is always lower.

## SCHOTTKY NOISE

Particle interaction affects the spectrum of beam Schottky noise. In the application to the beam with significant space charge, this problem was solved in Ref. [4] in a framework of the rigid-beam model. Comparison of this analytic solution with a particle tracking code and with real beam measurements was considered in Ref. [7]. In this section we apply the results obtained above to the problem of beam Schottky noise.

In the rigid-beam approximation, the spectral power of the transverse Schottky noise $\bar{x}^2(\nu)$ is [4]:

$$\bar{x}^2(\nu) = \frac{\varepsilon_x \beta_x}{N} \frac{P(\nu)}{|\varepsilon(\nu)|^2} \quad , \quad (11)$$

$$P(\nu) = \pi \int f J_x \, \delta(\Delta Q_l + \Delta Q_{sc} - \nu) d\Gamma$$

where $\beta_x$, $N$ are the beta-function at the pickup and the number of particles. Note that this result assumes the validity of the rigid-beam model, $|\nu - \Delta Q_{sc}| \gg \sigma_\nu \equiv \sigma(\Omega_i Q_i)/\Omega_0$. As above, it is true in vicinity of the coherent peak due to the high value of the tune separation. Since the model is not generally correct at the incoherent frequency range, $|\nu - \Delta Q_{sc}| \cong \sigma_\nu$, the above result is not justified there. However, the noise power (11) reaches its maximum at the coherent peak, where the model is valid, and, consequently, Eq. (11) can be used. Expansion of the denominator at $\nu = \nu_c + \Delta \nu$, $\Delta \nu \ll \Delta Q_{sep}$, leads to

$$\bar{x}^2(\nu) = \frac{\varepsilon_x \beta_x}{N} \frac{\kappa \Lambda(\nu_c)}{(\nu - \nu_c)^2 + (\Lambda(\nu_c) - \mathrm{Im}\,\Delta Q_c)^2}. \quad (12)$$

The form-factor

$$\kappa \equiv \frac{\langle \Delta Q_{sep} \rangle^2 P(\nu_c)}{\Lambda(\nu_c)},$$

introduced here, appears to be scarcely sensitive to the beam features, being always $\kappa \approx 1$. Indeed, let the tune spread be chromatic, with arbitrary momentum distribution. For KV distribution, $\kappa = 1$. For a Gaussian transverse distribution, a fit $\kappa = 0.9 + 0.02 \Delta Q_{sc}(0)/\sigma_{\nu p}$ is valid with accuracy of a few percent for any $\Delta Q_{sc}(0)/\sigma_{\nu p} \leq 20$. Integrating the noise power (12) over frequencies for the given Schottky band yields:

$$\langle \bar{x}^2 \rangle \equiv \int \bar{x}^2(\nu) \frac{d\nu}{\pi} = \kappa \frac{\varepsilon_x \beta_x}{N} \frac{1}{1 - \mathrm{Im}\,\Delta Q_c / \Lambda(\nu_c)} \quad (13)$$

Note that the integrated power (13) contains the coherent growth rate $\mathrm{Im}\,\Delta Q_c$ multiplied by a factor, extremely sensitive to the beam temperature, $\Lambda(\nu_c)^{-1}$. When the beam is being cooled, its total Schottky noise (13) almost does not change, being equal to its zero-impedance limit, until the very threshold of the instability, where it immediately jumps to infinity. That is why measuring the Schottky noise can hardly help to see a real part of impedance responsible for the coherent rate $\mathrm{Im}\,\Delta Q_c$: the rate is either invisible or fatal.

## SUMMARY

The applicability of the rigid-beam model is considered for the case when the space charge plays significant role in beam dynamics. The results prove that the stability diagrams obtained with this model are not valid for most of the complex plane of the coherent shift. However, the small area of its validity typically covers the entire area of practical interest. Based on the rigid-beam model, rather simple formulas for the Landau damping were calculated. These formulas are used for computation of the threshold space charge tune shift versus coherent growth time. Convenient analytical fits for the threshold lines are presented for round Gaussian beams.

The results obtained here do not undermine results of Refs [4-6], based on the analysis of the dispersion equation at the rigid-beam approximation, as soon as this approximation is valid. What is new in this paper is, first, a delimitation of validity area for the rigid-beam approximation, and, second, solutions of the dispersion equation are presented in more details, important for practical analysis.

At the end, the results are applied for the Schottky noise. For strong space charge case studied here, $|\nu - \Delta Q_{sc}| \gg \sigma_\nu$, the Schottky spectrum is dominated by narrow resonant peaks at coherent frequencies.

# APPENDIX

Here the solution (5) of the dispersion equation (4) is derived. The problem can be formulated as follows: for a given real eigenvalue $\nu$, the real and imaginary parts of the corresponding coherent tune shift have to be found. The solution can be sought as

$$\Delta Q_c = \nu - \delta Q + i\Lambda. \quad (A1)$$

A correction to the real part $\delta Q$ is found by expansion of the dispersion integral in (4) by the small lattice tune spread $\Delta Q_l / \Delta Q_{sep}$. Indeed, after substitution (A1) in the dispersion integral (4), and then addition and subtraction $\Delta Q_l$ in its numerator, the real part of the dispersion equation results in

$$\delta Q \int \frac{f_x J_x d\Gamma}{\Delta Q_l + \Delta Q_{sc} - \nu} = \int \frac{\Delta Q_l f_x J_x d\Gamma}{\Delta Q_l + \Delta Q_{sc} - \nu}. \quad (A2)$$

To first order in the small parameter $\Delta Q_l / \Delta Q_{sep}$ it leads to

$$\delta Q = \delta Q^{(1)} = -\langle \Delta Q_{sep} \rangle \int \frac{\Delta Q_l f_x J_x \, d\Gamma}{\Delta Q_{sep}};$$

$$\Delta Q_{sep} \equiv \operatorname{Re} \Delta Q_c - \Delta Q_{sc}; \; \langle \Delta Q_{sep} \rangle \equiv -\left( \int \frac{f_x J_x \, d\Gamma}{\Delta Q_{sep}} \right)^{-1}. \quad (A3)$$

This first-order result is sufficient for the octupole-related lattice tune shift, where $\delta Q^{(1)} \neq 0$, and the second-order term gives only a small correction to the Landau damping. For the chromatic tune spread though, the first-order correction vanishes, $\delta Q^{(1)} = 0$, and the second-order term has to be taken into account. The only non-zero second-order term comes from expansion of the denominator in the right-hand side of Eq. (A2) over the small parameter $\Delta Q_l / \Delta Q_{sep}$, resulting in

$$\delta Q = \delta Q^{(2)} = -\langle \Delta Q_{sep} \rangle \int \frac{\Delta Q_l^2 f_x J_x \, d\Gamma}{\Delta Q_{sep}^2}. \quad (A4)$$

If both octupole and chromatic tune spreads have to be taken into account, the former gives the first-order correction (A3), while the latter results in the second-order term (A4). Note also, that in the denominators of (A3, A4), the eigenvalue $\nu$ is substituted by the coherent shift $\operatorname{Re} \Delta Q_c$. Thus, the real part of the dispersion equation leads to

$$\nu = \operatorname{Re} \Delta Q_c + \delta Q, \quad (A5)$$

with $\delta Q$ as a sum of the two contributions (A3), (A4). An imaginary part of the dispersion integral is conventionally calculated using $\operatorname{Im}(1/(x-i0)) = \pi \delta(x)$, and immediately results in the damping rate $\Lambda$ in Eq. (5).